\documentclass[conference]{IEEEtran}
\IEEEoverridecommandlockouts

\usepackage{comment}
\usepackage{multirow}
\usepackage{booktabs}
\usepackage{enumitem}
\usepackage{hyperref}

\usepackage{cite}
\usepackage{amsmath,amssymb,amsfonts}
\usepackage{algorithmic}
\usepackage{graphicx}
\usepackage{textcomp}
\usepackage{xcolor}
\usepackage{orcidlink}
\def\BibTeX{{\rm B\kern-.05em{\sc i\kern-.025em b}\kern-.08em
    T\kern-.1667em\lower.7ex\hbox{E}\kern-.125emX}}
\begin{document}

\title{SG-Mamba: Sparse Graph-Guided Mamba for Audio-Visual Speech Enhancement}

  \author{
    \IEEEauthorblockN{Guo-Ruei Tseng\IEEEauthorrefmark{1} \orcidlink{0009-0002-8860-8629}, Hung-Shin Lee\IEEEauthorrefmark{3} \orcidlink{0000-0001-7044-9434}, Hsin-Min Wang\IEEEauthorrefmark{2} \orcidlink{0000-0003-3599-5071}, and Berlin Chen\IEEEauthorrefmark{1} \orcidlink{0000-0003-0693-8932}}

    \IEEEauthorblockA{
        \IEEEauthorrefmark{1}Dept. of Computer Science and Information Engineering, National Taiwan Normal University, Taiwan\\
        \IEEEauthorrefmark{2}Inst. of Information Science, Academia Sinica, Taiwan\\
        \IEEEauthorrefmark{3}Grad. Inst. of AI Interdisciplinary Applied Technology, National Taiwan Normal University, Taiwan
    }
  }

\maketitle

\begin{abstract}
Lightweight audio-visual speech enhancement (AVSE) models face a critical trade-off between computational efficiency and cross-modal alignment accuracy. While simple concatenation lacks relational expressiveness, dense cross-attention incurs computational overhead and is prone to unreliable cross-modal correspondence under strong acoustic interference. We propose Sparse Graph-Guided Mamba (SG-Mamba), a lightweight AVSE framework that integrates a sparse heterogeneous graph with a linear-complexity Mamba backbone. The graph explicitly models modality-specific relations through content-adaptive attention and cross-frame audio-visual connections, while Mamba captures long-range temporal context. We further introduce an audio skip connection to preserve spectral detail without sacrificing noise suppression. Evaluated on LRS3, SG-Mamba achieves competitive or superior performance against strong lightweight baselines and reaches 13.091 dB SI-SDR under noise-only condition. It also remains robust in cluttered multi-speaker conditions with a competitive cost of 3.45 G MACs (or 6.90 G FLOPs). Results on VoxCeleb2 further suggest that explicit structural priors improve robustness, generalizability, and computational efficiency in lightweight AVSE.
\end{abstract}

\begin{IEEEkeywords}
audio-visual speech enhancement, graph-guided multimodal fusion, graph neural networks, state space models, Mamba
\end{IEEEkeywords}

\section{Introduction}
Speech enhancement (SE) is a core component in applications such as smartphones and hearing aids, where the objective is to recover clean speech from noisy observations \cite{loizou2007speech, wang2018supervised}.
Compared with audio-only systems, audio-visual speech enhancement (AVSE) improves robustness by leveraging visual cues such as lip motion \cite{afouras2018conversation, michelsanti2021overview, hou2018audio}.
Recently, Real-Time Audio-Visual Speech Enhancement Using Pre-trained Visual Representations (RAVEN) \cite{ma25c_interspeech} demonstrated strong performance by leveraging robust pre-trained visual front-ends \cite{shi2022robust, ma2022visual, tao2021someone, wang2024loconet}. 
To further improve temporal sequence modeling while maintaining computational efficiency, Mamba \cite{zhang2025mamba, qian2025sav, chao2024investigation, wang2025mamba} provides linear-complexity sequence modeling while maintaining competitive long-range dependency modeling, making it an attractive backbone for lightweight AVSE.

Despite these advances, fusion design in AVSE still faces a trade-off between computational efficiency (e.g., latency and parameter count) and the expressive capacity required for effective cross-modal relational modeling \cite{xu2023multi}.
Methods with strong enhancement performance, such as diffusion \cite{richter2023audio, ayilo2025diffusion, chou2024av2wav} and flow-matching models \cite{jung24b_interspeech}, typically require expensive sampling or parameter-heavy backbones.
Transformer-based methods \cite{mira2023voce, wahab2024multi, sajid25_avsec} achieve strong performance through expressive attention-based backbones, but their deep architectures remain costly for practical deployment. 
In contrast, efficient architectures often fall into a ``Concatenation Trap,'' relying on simple concatenation, feature-wise modulation (FiLM) \cite{ahmed2025av}, or linear attention without explicit structural constraints.
These designs fuse heterogeneous modalities in a shared feature space without explicitly modeling their structural correspondence, forcing the backbone to infer cross-modal relations implicitly. 
Alternatively, increasing the interaction capacity through dense global cross-attention \cite{saleem2025viseme} partially alleviates this limitation; however, without explicit structural constraints, these global interactions remain susceptible to unreliable cross-modal correspondence under severe acoustic interference \cite{liu2025alignvsr}. 
Therefore, existing AVSE systems lack a structurally bounded fusion mechanism that simultaneously accommodates local audio-visual micro-asynchrony \cite{chandrasekaran2009natural, schwartz2014no} while preventing erroneous alignments from propagating globally. 

\begin{figure*}
\centering
\includegraphics[width=\textwidth]{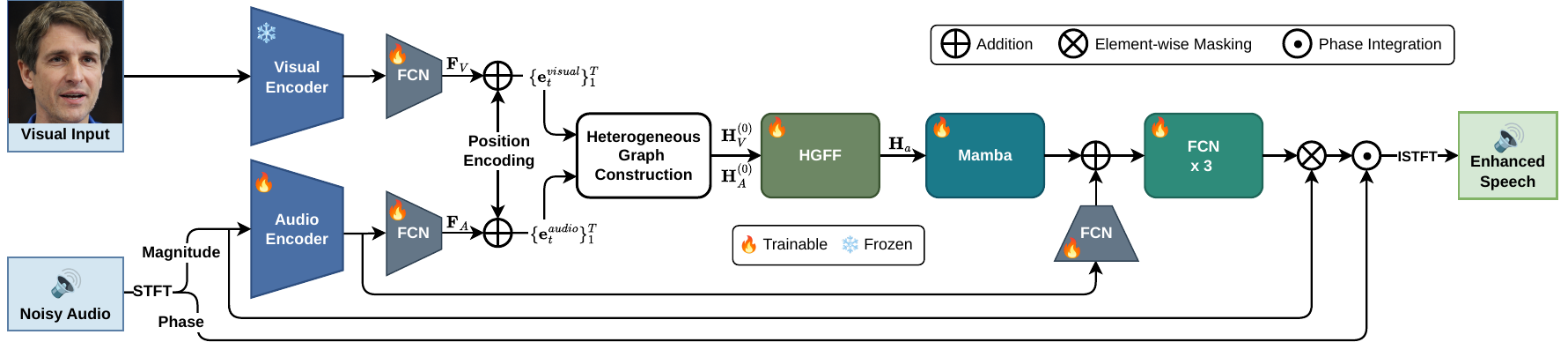}
\vspace{-15pt}
\caption{Architecture of the proposed SG-Mamba.
Modality-specific encoders extract audio and visual features ($\mathbf{F}_A, \mathbf{F}_V$), which are position-encoded and fed into the Heterogeneous Graph Construction module to yield node feature matrices $\mathbf{H}^{(0)}_A$ and $\mathbf{H}^{(0)}_V$.
A HGFF module then performs content-adaptive cross-modal alignment, outputting $\mathbf{H}_a$ to a unidirectional Mamba block for global temporal modeling.
Finally, 3 stacked FCN layers generate a mask for magnitude refinement. 
The refined magnitude is then integrated with the original noisy phase and reconstructed via ISTFT.}
\vspace{-15pt}
\label{fig1}
\end{figure*}

Motivated by these limitations and inspired by the success of graph neural networks in multimodal fusion \cite{chen2020hgmf} and speech enhancement \cite{chau2024novel}, we propose Sparse Graph-Guided Mamba (SG-Mamba) \footnote{The source code is available at: \url{https://github.com/0u5gary/SG-Mamba}}, a lightweight framework that employs explicit sparse graph-guided cross-modal interaction. 
Our main contributions are summarized as follows:

\begin{enumerate}[noitemsep,leftmargin=*]

\item \textbf{Heterogeneous Graph for Fine-Grained Fusion:}
We replace dense quadratic attention with a sparse heterogeneous graph processed via dynamic, content-adaptive attention. This explicitly models modality-specific interactions under varying acoustic conditions, striking a practical balance between the weak structural prior of concatenation and the high computational cost of full attention.

\item \textbf{Explicit Cross-Frame Modeling:} 
Instead of relying entirely on the backbone to implicitly infer temporal offsets, we explicitly introduce cross-frame edges connecting neighboring visual frames to the corresponding audio nodes. 
The resulting graph propagation enables flexible local alignment before temporal sequence modeling, improving robustness under temporal mismatch conditions.

\item \textbf{Synergistic SG-Mamba Backbone:} 
We combine a U-Net-structured dynamic graph attention network (GAT) \cite{brody2022how} with a state-space architecture (Mamba). 
By dedicating the graph-based U-Net to hierarchical cross-modal relational aggregation and the state-space model to long-range utterance-level temporal modeling, this targeted architectural division effectively mitigates severe acoustic interference without inflating computational costs.
\end{enumerate}

\section{Methodology}
Our framework extends RAVEN \cite{ma25c_interspeech} by introducing a structurally graph-guided fusion module and a Mamba-based temporal modeling backbone.
As illustrated in Fig. 1, SG-Mamba adopts a modular four-stage design for lightweight offline AVSE, which consists of: (1) Feature Extraction and Node Embedding, (2) Heterogeneous Graph Construction, (3) Hierarchical Graph-Structured Feature Fusion, and (4) Mamba Modeling and Speech Reconstruction.

\subsection{Feature Extraction and Node Embedding}
Given a noisy audio stream and a synchronized visual stream, we first apply a Short-Time Fourier Transform (STFT) to obtain the magnitude and phase spectrograms, followed by power compression $p$ of the magnitude.
We feed the magnitude spectrogram into an Audio Encoder (five CNN layers, each followed by BatchNorm and ReLU, with a temporal receptive field of 5), followed by a Fully Connected Network (FCN) layer for dimensionality reduction, yielding audio features $\mathbf{F}_A \in \mathbb{R}^{T \times D}$, where $T$ denotes the total number of frames and $D$ denotes the feature dimension.
We retain the phase component for final waveform reconstruction.
Similarly, we pass visual frames through a pre-trained visual encoder, then temporally upsample them to match the audio frame rate, and apply an FCN to obtain lip-reading visual features $\mathbf{F}_V \in \mathbb{R}^{T \times D}$.

Because graph message passing does not inherently encode temporal order, we add Sinusoidal Positional Encoding ($PE$) \cite{vaswani2017attention} to both modalities to preserve temporal order.
We obtain the final node embeddings as $\mathbf{e}_t^{m} = \mathbf{f}_t^{m} + PE_t, \quad m \in \{\operatorname{audio}, \operatorname{visual}\}$, where $\mathbf{e}_t^{m}$ denotes the embedding vector for modality $m$ at frame $t$ and serves as the input to Heterogeneous Graph Construction.

\begin{figure}
\centering
\includegraphics[width=0.8\columnwidth]{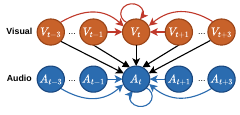}
\vspace{-10pt}
\caption{The proposed heterogeneous graph centered at the current frame $t$.
Red and blue arrows denote intra-modal temporal connections within each modality, while black arrows indicate unidirectional cross-modal interactions from the local visual window $V_{t \pm 3}$ to the audio node $A_t$.
}
\vspace{-15pt}
\label{fig2}
\end{figure}

\subsection{Heterogeneous Graph Construction}
To explicitly capture inter-modal interactions and accommodate dynamic micro-asynchrony, we construct a heterogeneous graph $\mathcal{G} = (\mathcal{V}, \mathcal{E})$. 
The node set $\mathcal{V}$ is partitioned into an audio subset $\mathcal{V}_a = \{A_1, \dots, A_t, \dots, A_T\}$ and a visual subset $\mathcal{V}_v = \{V_1, \dots, V_t, \dots, V_T\}$. 
Each node is initialized with embedding $\mathbf{e}_t^{m}$, yielding the audio and visual node feature matrices $\mathbf{H}^{(0)}_A$ and $\mathbf{H}^{(0)}_V$, which are fed separately into the Hierarchical Graph-Structured Feature Fusion (HGFF) module. 

As shown in Fig. \ref{fig2}, all nodes retain self-loops ($A_t \to A_t$, $V_t \to V_t$) to preserve frame-level intrinsic features during message passing. 
To model intra-modal temporal continuity, we connect neighboring frames from $t-3$ to $t+3$ to the current node at frame $t$ within each modality. 
For cross-modal interactions, audio node $A_t$ establish directed edges from a local visual window $V_{t \pm 3}$ (approximately $\pm 30$ ms).   
Crucially, these cross-modal edges are strictly unidirectional ($V_{t \pm 3} \to A_t$), as allowing noisy acoustic features to propagate into the visual stream could destabilize its cleaner representations.
Moreover, restricting aggregation to a local $\pm 3$-frame window reduces computational overhead. 
This design deliberately trades long-range dense alignment for local stability, ensuring the cross-modal graph remains lightweight while effectively capturing immediate phoneme-viseme correlations. 
Formally, we define the temporal neighbor window at frame $t$ as $\mathcal{N}_t = \{t+k \mid k \in [-3, 3],\ 0 \le t+k < T\}$, which is applied to audio-audio (AA), visual-audio (VA), and visual-visual (VV) interaction edge types.
The resulting graph connectivity is denoted by the adjacency matrix $\mathbf{A}_\mathcal{G}$. 
In practice, $\mathbf{A}_\mathcal{G}$ is implemented as two sparse neighbor index tables of shape $T \times k$ (where $k = 2 \times 3 + 1 = 7$), one shared across AA and VV interactions and one for VA, avoiding materialization of the full $2T \times 2T$ matrix across the batch dimension. 

\subsection{Hierarchical Graph-Structured Feature Fusion}
\begin{figure}
\centering
\includegraphics[width=0.95\columnwidth]{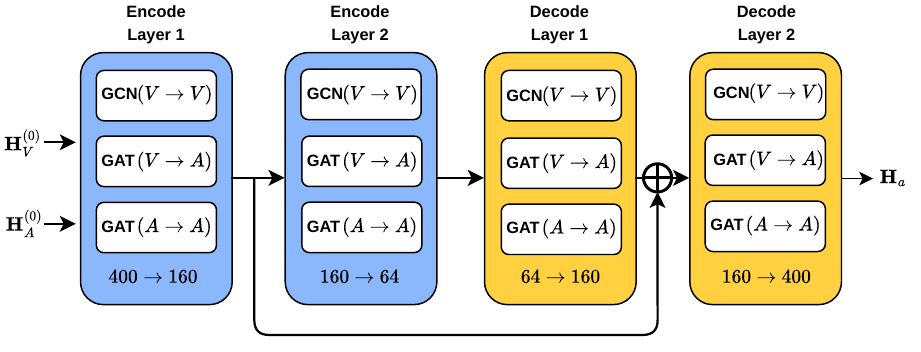}
\vspace{-10pt}
\caption{Detailed architecture of the HGFF module, featuring modality-specific graph operations within a symmetrical U-Net backbone.}
\vspace{-15pt}
\label{fig3}
\end{figure}

As illustrated in Fig. \ref{fig3}, we employ the HGFF module to perform content-adaptive, fine-grained cross-modal structural alignment. 
Within each layer $l$ in HGFF, the three edge types defined in $\mathbf{A}_{\mathcal{G}}$ are handled differently according to their respective characteristics to yield the layer's audio and visual features, denoted as $\mathbf{H}^{(l)}_A$ and $\mathbf{H}^{(l)}_V$. 
Let $[\mathbf{m}_{(\cdot)}]_t \in \mathbb{R}^{D}$ denote the aggregated message vector at frame $t$ for interaction type $(\cdot) \in \{AA, VA, VV\}$.
For interaction types $AA$ and $VA$, the resulting message vector is aggregated via a GAT layer:
\begin{equation}
  [\mathbf{m}_{(\cdot)}]_t = \sum_{j \in \mathcal{N}_t} \alpha_{tj} 
  \mathbf{W}_{(\cdot)}\mathbf{h}^{(l)}_j,
\end{equation}
where $\mathbf{h}^{(l)}_j$ denotes the features of neighbor $j$ drawn from $\mathbf{H}^{(l)}_A$ for AA and $\mathbf{H}^{(l)}_V$ for VA, and $\mathbf{W}_{(\cdot)}$ is a linear projection matrix. 
Specifically, to compute the attention weight $\alpha_{tj}$, which weights the contribution from neighbor $j$ to $t$, we adopt the dynamic graph attention mechanism \cite{brody2022how}. 
This approach avoids the static attention degradation present in standard GAT by applying a joint nonlinear transformation (LeakyReLU) to the query-key pair before projecting it into a scalar score via a learnable vector $\mathbf{a}$. 
Subsequently, a learnable temperature parameter $\tau$ is applied before softmax normalization to prevent degenerate attention distributions and stabilize training. 
All parameters and computations, including $\mathbf{a}$, $\tau$, and the resulting $\alpha_{tj}$, are maintained and performed independently for AA and VA to capture the distinct dynamics of each interaction type. 
For the remaining interaction type $VV$, a Graph Convolutional Network (GCN) mean aggregation is adopted:
\begin{equation}
  [\mathbf{m}_{(\cdot)}]_t = \mathbf{W}_{(\cdot)} \left( \frac{1}{|\mathcal{N}_t|} \sum_{j \in \mathcal{N}_t} \mathbf{h}^{(l)}_j \right),
\end{equation} 
where $\mathbf{h}^{(l)}_j$ is drawn from $\mathbf{H}^{(l)}_V$.
This uniform aggregation stabilizes visual anchor representations without attention overhead, reflecting the temporal redundancy in upsampled lip features, where highly correlated frames make averaging sufficient.  

Stacking $[\mathbf{m}_{(\cdot)}]_t$ across all $T$ frames yields the message matrix $\mathbf{M}_{(\cdot)} \in \mathbb{R}^{T \times D}$.
The audio and visual node feature matrices are then updated via projected residual connections followed by layer normalization (LN):
\begin{equation}
    \mathbf{H}^{(l+1)}_A \leftarrow \text{LN}\!\left(
    \mathbf{W}_A\,\mathbf{H}^{(l)}_A + 
    \mathbf{M}_{AA} + \mathbf{M}_{VA}\right),
\end{equation}
\begin{equation}
    \mathbf{H}^{(l+1)}_V \leftarrow \text{LN}\!\left(
    \mathbf{W}_V\,\mathbf{H}^{(l)}_V + 
    \mathbf{M}_{VV}\right),
\end{equation}
where $\mathbf{W}_A$ and $\mathbf{W}_V$ are residual projections applied only when input and output dimensions differ. 
These HGFF layers are stacked into a 4-layer U-Net backbone, compressing dimensions at the encoder and symmetrically expanding at the decoder, with skip additions introduced before the final decoder block to preserve lower-level fine-grained cues, resulting in an effective receptive field of $\pm 12$ frames ($\pm 120$ ms). 
Furthermore, this architecture provides inherent structural constraints and noise-robust multi-scale feature representations, ensuring stable multi-modal integration even under severe acoustic degradation. 
After the final decoder layer, we retain only the aggregated audio node representations $\mathbf{H}_{a} \in \mathbb{R}^{T \times D}$, which are passed to the subsequent Mamba block.

Unlike uniform 1D convolutions that mix modalities indiscriminately, or Transformer-based cross-attention that incurs quadratic $\mathcal{O}(T^2)$ complexity, our approach encodes explicit cross-modal structural priors through the sparse asymmetric $\mathbf{A}_\mathcal{G}$. 
By routing cross-modal information dynamically via GAT while using lightweight GCN mean pooling for highly redundant visual features, this sparse topology bridges the audio-visual semantic gap while preserving the flexibility of dynamic representation learning within a tightly controlled computational budget. 

\subsection{Mamba Modeling and Speech Reconstruction}
To capture global temporal dependencies, we feed the aggregated audio features $\mathbf{H}_{a}$ into a unidirectional Mamba block. 
Since local bidirectional context has already been explicitly aggregated by the HGFF module, the subsequent Mamba is only responsible for modeling residual long-range temporal dependencies. 
Therefore, we adopt a unidirectional scan to avoid redundant temporal modeling while maintaining a lightweight backend.
Specifically, Mamba performs selective scanning with input-dependent parameters $(\Delta_t, \mathbf{B}_t, \mathbf{C}_t)$.
For the $t$-th frame $\mathbf{x}_t \in \mathbb{R}^D$ of $\mathbf{H}_{a}$, the discrete state update is:
\begin{equation}
\mathbf{h}_t = \mathbf{\bar{A}}_t \mathbf{h}_{t-1} + \mathbf{\bar{B}}_t \mathbf{x}_t, \quad \mathbf{y}_t = \mathbf{C}_t \mathbf{h}_t,
\end{equation}
where the discretized parameters $\mathbf{\bar{A}}_t = \exp(\Delta_t \mathbf{A})$ and $\mathbf{\bar{B}}_t \approx \Delta_t \mathbf{B}_t$ (derived from the base transition $\mathbf{A}$) dynamically gate information flow and improve robustness.

To align latent dimensions and compensate for Mamba compression, we add the Mamba output with the skip connection from $\mathbf{F}_A$, and then apply LN to the summed representation.
Three stacked FCN layers (each with ReLU activation) then serve as a non-linear fusion head, reconciling Mamba's structural consistency with fine spectral details. 
Subsequently, a sigmoid activation generates a multiplicative mask $\mathbf{M} \in [0, 1]^{T \times D}$.
We apply $\mathbf{M}$ to the power-compressed magnitude, invert the compression $1/p$, combine with the noisy phase, and reconstruct the enhanced waveform via Inverse Short-Time Fourier Transform (ISTFT).

For training, following RAVEN \cite{ma25c_interspeech}, we adopt a composite spectral loss that combines complex-domain and magnitude reconstruction terms to jointly supervise phase consistency and spectral fidelity:
\begin{equation}
\mathcal{L}_{total}(\mathbf{\hat{S}}, \mathbf{S}) = {|| \mathbf{\hat{S}} - \mathbf{S} ||}_2 + {|| |\mathbf{\hat{S}}| - |\mathbf{S}| ||}_2,
\end{equation}
where $\mathbf{\hat{S}}$ and $\mathbf{S}$ denote the estimated and clean power-compressed spectrograms, respectively.

\section{Experimental Setup}
\subsection{Dataset and Evaluation}
We used 433 hours of audio-visual data from LRS3 \cite{afouras2018lrs3} to build target speech, visual cues, and interfering speech samples, together with 180 hours of background noise and reverberation data from the DNS Challenge \cite{reddy2020interspeech}. 
Following the official LRS3 split, the pretrain, trainval, and test sets were used for training, validation, and evaluation respectively, ensuring no speaker overlap across splits.
To simulate complex acoustic environments, we dynamically mixed target speech with either LRS3 interferers or diverse DNS noise at SNRs uniformly sampled from -10 dB to 10 dB, covering both low- and high-interference conditions.
All audio-visual segments were truncated or zero-padded to a fixed duration of 5 seconds for consistent model input length.
To further evaluate cross-dataset generalizability, we applied the LRS3-trained models directly to the official VoxCeleb2 test set \cite{chung18b_interspeech} without fine-tuning, using the same mixing protocol and SNR range.
We evaluated performance using three standard metrics: PESQ \cite{rix2001perceptual} for perceptual quality, ESTOI \cite{jensen2016algorithm} for intelligibility, and SI-SDR \cite{le2019sdr} for signal-level distortion reduction.
Among these, SI-SDR is used as the sole metric in Secs. \ref{Ablation} and \ref{cross} to highlight performance differences. 

\subsection{Implementation Setup}
The audio stream was resampled to 16 kHz and converted into 257-dimensional power-compressed magnitude spectrograms (25 ms Hann window, 10 ms hop size, 512-point FFT, 0.3 compression rate), while the video stream provided $96 \times 96$ lip ROIs.
Following RAVEN \cite{ma25c_interspeech}, we extracted visual features using the frozen pre-trained visual front-end from AVHuBERT \cite{shi2022robust} and TalkNet \cite{tao2021someone}, whose outputs were concatenated into 1,280-dimensional embeddings. 
Since the visual front-end is frozen and identical across all compared methods, the term ``lightweight'' refers to the trainable fusion and enhancement backend, whose complexity is reported for fair architectural comparison. 
Comparisons are limited to reproducible lightweight backbones under a unified experimental protocol. 
The backend consisted of the proposed HGFF backbone with single-head GAT (compressing from $D=400$ to a bottleneck of 64 and symmetrically expanding to 400) followed by a unidirectional Mamba block ($d_\text{state}=64$, $d_\text{conv}=4$, expansion factor $=2$). 
All compared models were trained with the Adam optimizer at an initial learning rate of $10^{-5}$. 
To ensure fair comparison, no architecture-specific hyperparameter tuning was performed. 
A dropout rate of 0.1 was applied to all attention-based and graph aggregation layers.
To stabilize training, a dynamic learning rate scheduling strategy was employed, which halved the learning rate when the validation loss plateaued.
We used a micro-batch size of 16 with gradient accumulation, yielding an effective batch size of 128.

\subsection{Baseline Configuration}
To assess the framework under lightweight constraints, all baselines share the same audio and visual encoders and operate at a hidden dimension of $D = 400$. 
RAVEN serves as the primary baseline, with two modifications applied for comparability. First, the concatenated audio-visual features are projected to $D = 400$ via an FCN before being fed into the LSTM, rather than passing the full high-dimensional representation directly as in the original implementation. Second, an audio skip connection with layer normalization is added before the output FCN layers to match the reconstruction pipeline of the proposed method. 
The Mamba baseline follows the same design principle as the modified RAVEN, replacing the LSTM with a unidirectional Mamba block under otherwise identical configurations. 
The Bi-Mamba ablation baseline extends the Mamba baseline with a bidirectional design, where forward and backward scans are performed independently and combined via element-wise addition prior to the residual connection. 
For the remaining ablation baselines, audio and visual features are independently projected to $D = 400$ before fusion: FiLM-Mamba applies feature-wise linear modulation to inject visual conditioning into the audio stream, Linear-Mamba employs a linear attention mechanism \cite{katharopoulos2020transformers} where audio attends to visual features, and Cross-Mamba uses a 4-head cross-attention module where audio queries attend to visual keys and values. 
All four ablation baselines also incorporate the same audio skip connection as the RAVEN and Mamba baselines. 
SG-RAVEN adopts the same heterogeneous graph front-end as SG-Mamba but substitutes the Mamba block with LSTM, serving as a direct backbone comparison under identical fusion conditions.

\begin{table}[t]
\centering
\footnotesize
\caption{Performance comparison under noise-only scenario with mixed SNR condition from [-10 dB, 10 dB].
}
\vspace{-5pt}
\label{tab:noise_results}
\setlength{\tabcolsep}{3pt}
\begin{tabular}{lccccc}
\toprule
\textbf{Method} & \textbf{PESQ} & \textbf{SI-SDR} & \textbf{ESTOI} & \textbf{Para. ($\downarrow$)} & \textbf{MACs ($\downarrow$)} \\
\midrule
Noisy & 1.841 & 2.551 & 0.812 & -- & --\\
\midrule
RAVEN (Base) \cite{ma25c_interspeech} & 2.300 & 12.648 & 0.837 & \underline{4.7 M} & \underline{3.19 G} \\
Mamba & \textbf{2.305} & 12.463 & \underline{0.840} & \textbf{4.5 M} & \textbf{3.16 G} \\
\midrule
SG-RAVEN & 2.286 & \underline{12.857} & 0.838 & 5.4 M & 3.47 G \\
\textbf{SG-Mamba} & \underline{2.302} & \textbf{13.091} & \textbf{0.842} & 5.3 M & 3.45 G \\
\bottomrule
\end{tabular}
\vspace{-15pt}
\end{table}

\begin{table*}[t!]
\centering
\footnotesize
\caption{Performance comparison under \textbf{1-Interferer} scenario across different SNR levels. The interfering speaker is not visible.}
\vspace{-5pt}
\label{tab:interferer_1}
\setlength{\tabcolsep}{8.5pt}
\begin{tabular}{lccc|ccc|ccc}
\toprule
\multirow{2}{*}{\textbf{Method}} & \multicolumn{3}{c}{\textbf{0 dB}} & \multicolumn{3}{c}{\textbf{5 dB}} & \multicolumn{3}{c}{\textbf{10 dB}} \\ 
& \textbf{PESQ} & \textbf{SI-SDR} & \textbf{ESTOI} & \textbf{PESQ} & \textbf{SI-SDR} & \textbf{ESTOI} & \textbf{PESQ} & \textbf{SI-SDR} & \textbf{ESTOI} \\
\midrule
Noisy & 1.314 & 0.941 & 0.627 & 1.553 & 5.942 & 0.732 & 1.956 & \textbf{10.942} & \underline{0.824} \\
\midrule
RAVEN (Base) \cite{ma25c_interspeech} & 1.493 & 2.176 & 0.636 & \underline{1.763} & 5.871 & 0.738 & \underline{2.138} & 8.773 & 0.822 \\
Mamba & \underline{1.496} & \underline{2.474} & \underline{0.644} & 1.761 & 6.045 & \underline{0.742} & 2.124 & 8.640 & 0.822 \\
\midrule
SG-RAVEN & 1.485 & 2.388 & 0.637 & 1.750 & \underline{6.124} & 0.739 & 2.127 & 9.054 & 0.823 \\
\textbf{SG-Mamba (Ours)} & \textbf{1.509} & \textbf{2.752} & \textbf{0.647} & \textbf{1.775} & \textbf{6.229} & \textbf{0.745} & \textbf{2.143} & \underline{9.096} & \textbf{0.827} \\
\bottomrule
\end{tabular}
\vspace{-13pt}
\end{table*}

\begin{table*}[t!]
\centering
\footnotesize
\caption{Performance comparison under \textbf{3-Interferer} scenario across different SNR levels. The interfering speaker is not visible.}
\vspace{-5pt}
\label{tab:interferer_3}
\setlength{\tabcolsep}{8.5pt}
\begin{tabular}{lccc|ccc|ccc}
\toprule
\multirow{2}{*}{\textbf{Method}} & \multicolumn{3}{c}{\textbf{0 dB}} & \multicolumn{3}{c}{\textbf{5 dB}} & \multicolumn{3}{c}{\textbf{10 dB}} \\ 
& \textbf{PESQ} & \textbf{SI-SDR} & \textbf{ESTOI} & \textbf{PESQ} & \textbf{SI-SDR} & \textbf{ESTOI} & \textbf{PESQ} & \textbf{SI-SDR} & \textbf{ESTOI} \\
\midrule
Noisy & 1.200 & 1.064 & 0.526 & 1.406 & \textbf{6.064} & 0.665 & 1.794 & \textbf{11.063} & 0.786 \\
\midrule
RAVEN (Base) \cite{ma25c_interspeech} & 1.392 & 1.841 & 0.558 & 1.665 & 5.485 & 0.692 & \underline{2.048} & 8.511 & 0.799 \\
Mamba & \underline{1.401} & \underline{2.293} & \underline{0.565} & \underline{1.672} & 5.716 & \underline{0.696} & 2.044 & 8.414 & 0.800 \\
\midrule
SG-RAVEN & 1.385 & 2.145 & 0.556 & 1.654 & 5.808 & 0.694 & 2.035 & 8.842 & \underline{0.801} \\
\textbf{SG-Mamba (Ours)} & \textbf{1.403} & \textbf{2.372} & \textbf{0.567} & \textbf{1.676} & \underline{6.011} & \textbf{0.703} & \textbf{2.056} & \underline{9.190} & \textbf{0.810} \\ \bottomrule
\end{tabular}
\vspace{-15pt}
\end{table*}

\section{Results}
\subsection{Performance on Non-Speech Noise}
As shown in Table~\ref{tab:noise_results}, SG-Mamba achieved the highest SI-SDR of 13.091 dB, while maintaining stable PESQ and ESTOI comparable to the strongest baselines, indicating that the graph-guided fusion improves signal-level separation without compromising perceptual quality or intelligibility.
Compared with the strongest baselines, SG-Mamba improved SI-SDR by 0.443 dB over RAVEN and 0.628 dB over Mamba, confirming that explicit structural guidance at the fusion stage yields consistent signal-level gains under non-speech noise. 
Notably, SG-RAVEN also outperformed RAVEN by 0.209 dB in SI-SDR, suggesting that the heterogeneous graph fusion provides complementary structural cues that benefit both backbone types under diverse noise conditions. 
Nevertheless, the performance gap between SG-Mamba and SG-RAVEN (0.234 dB) indicates that Mamba's selective state-space mechanism integrates the graph-aggregated features more effectively than LSTM-based gating, particularly in leveraging long-range temporal context for noise suppression. 
These improvements were achieved with 5.3 M parameters and 3.45 G MACs, demonstrating a favorable performance-complexity trade-off.

\subsection{Performance on Interfering Speakers}
To evaluate robustness against off-screen interfering voices---mimicking real-world scenarios where background chatter mismatches visible lip movements---we tested under 1- and 3-interferer conditions across 0, 5, and 10 dB SNRs. 
As expected, all methods showed reduced SI-SDR at 10 dB relative to the unprocessed noisy input, reflecting that mask-based enhancement on relatively clean speech can introduce reconstruction artifacts that outweigh the suppressed interference. 
Despite this, SG-Mamba consistently achieved the best or competitive performance across all SNR levels and interferer counts, demonstrating its strong robustness against various degrees of acoustic interference.

\subsubsection{One Interfering Speaker}
As shown in Table~\ref{tab:interferer_1}, SG-Mamba achieved the best performance on all metrics across all SNR levels, confirming that the graph-guided fusion provides consistent gains over the RAVEN and Mamba baselines under single-interferer conditions. 
SG-Mamba outperformed Mamba in SI-SDR by 0.278 dB at 0 dB and 0.184 dB at 5 dB. 
SG-RAVEN similarly improved over RAVEN, suggesting that the heterogeneous graph front-end consistently provides structural benefits. 
Crucially, these gains became more pronounced under lower-SNR conditions: at 0 dB, SG-Mamba achieved an SI-SDR of 2.752 dB, outperforming RAVEN by 0.576 dB, whereas the margin narrowed at 5 dB (0.358 dB) and further at 10 dB, suggesting that explicit cross-modal structural priors are most effective when acoustic interference is most severe and visual anchoring is most critical.
Moreover, SG-Mamba also attained the highest PESQ and ESTOI, indicating that the structural gains extended to perceptual quality and intelligibility. 
Such advantages are likely attributable to the graph's explicit cross-modal routing, which enables the audio stream to selectively attend to visual cues as a stable anchor when a single competing voice disrupts the acoustic scene.

\begin{table*}[t!]
\centering
\footnotesize
\caption{Performance comparison under different fusion methods on SI-SDR.}
\vspace{-5pt}
\label{tab:Ablation}
\setlength{\tabcolsep}{8.5pt}
\begin{tabular}{lc|ccc|ccc|cc}
\toprule
\multirow{2}{*}{\textbf{Method}} & \multicolumn{1}{c}{\textbf{Noise Only}} & \multicolumn{3}{c}{\textbf{1-Interferer}} & \multicolumn{3}{c}{\textbf{3-Interferer}} & \multirow{2}{*}{\textbf{Para. ($\downarrow$)}} & \multirow{2}{*}{\textbf{MACs ($\downarrow$)}} \\ 
& \textbf{Mixed} & \textbf{0 dB} & \textbf{5 dB} & \textbf{10 dB} & \textbf{0 dB} & \textbf{5 dB} & \textbf{10 dB} & & \\
\midrule
Bi-Mamba & \underline{12.939} & 2.607 & 5.517 & 7.652 & 2.210 & 5.306 & 7.769 & 5.7 M & 3.27 G \\
Film-Mamba & 12.578 & \underline{2.641} & \underline{6.047} & 8.543 & \underline{2.305} & 5.548 & 8.105 & \textbf{4.9 M} & \textbf{3.20 G}\\
Linear-Mamba & 12.461 & 0.295 & 5.504 & 9.002 & 1.205 & 5.861 & 8.924 & \underline{5.2 M} & \underline{3.26 G}\\
Cross-Mamba & 12.583 & 0.198 & 5.466 & \underline{9.023} & 1.055 & \underline{5.868} & \underline{9.035} & \underline{5.2 M} & 3.27 G\\
\textbf{SG-Mamba (Ours)} & \textbf{13.091} & \textbf{2.752} & \textbf{6.229} & \textbf{9.096} & \textbf{2.372} & \textbf{6.011} & \textbf{9.190} & 5.3 M & 3.45 G\\
\bottomrule
\end{tabular}
\vspace{-20pt}
\end{table*}

\subsubsection{Three Interfering Speakers}
As shown in Table~\ref{tab:interferer_3}, SG-Mamba maintained its advantage under the most challenging three-interferer scenario, achieving the best PESQ and ESTOI across all SNR levels and competitive SI-SDR throughout, demonstrating that the graph-guided fusion remains effective even under dense speech-to-speech mixtures.
Compared to their non-graph counterparts, both SG-Mamba and SG-RAVEN yielded consistent SI-SDR improvements, suggesting that the heterogeneous graph front-end continues to provide structural benefits even as interferers increase.
Notably, compared with the single-interferer condition, the absolute SI-SDR margins between SG-Mamba and the non-graph baselines narrowed across all SNR levels, which we attributed to the fixed graph topology: 
while content-adaptive attention weights remain dynamic, the pre-defined sparse local connectivity constrains the receptive field of cross-modal aggregation, making it less capable of suppressing interference patterns that span a broader or more irregular acoustic context under dense mixtures. 
Nevertheless, SG-Mamba exhibited no performance collapse across any condition, and its consistent lead in PESQ and ESTOI suggests that the visual anchor effectively preserved perceptual quality and intelligibility even when signal-level separation became more challenging. 

\subsection{Ablation Studies} \label{Ablation}
\subsubsection{Impact of Fusion Strategy}
To evaluate our solution to the unstructured ``Concatenation Trap,'' we compared multiple fusion mechanisms, as illustrated in Table~\ref{tab:Ablation}. 
Despite carrying the highest parameter count (5.7 M), Bi-Mamba does not yield consistent gains under interferer conditions, possibly because the backward scan introduces conflicting future-frame context that exacerbates cross-source contamination. 
While stable across all conditions, FiLM-Mamba's feature-wise modulation lacks the structural expressiveness needed for effective cross-modal integration.  
Attention-based methods achieved competitive SI-SDR in the noise-only and 10 dB conditions but degraded noticeably at 0 dB, accompanied by a PESQ degradation of approximately 0.2 across all interference conditions.
We hypothesize that this degradation arises from negative transfer in unconstrained cross-modal fusion: without explicit structural boundaries, dense attention can erroneously align interfering acoustic features with the target's visual lip cues, causing the mask to suppress target speech. 
In contrast, SG-Mamba alleviates this by restricting cross-modal interactions to a sparse local window, bounding the propagation of such misalignment. 
Under a similar parameter budget, SG-Mamba achieved the best SI-SDR across all conditions while remaining stable under both interferer scenarios, suggesting a stronger robustness-fidelity balance than either feature modulation or dense attention under lightweight constraints.

\begin{table}[t]
\centering
\footnotesize
\caption{Ablation on graph hyperparameters $(k, \delta)$.}
\vspace{-7pt}
\label{tab:ablation_hop}
\setlength{\tabcolsep}{3pt}
\begin{tabular}{lccc}
\toprule
$(K, \delta)$ & \textbf{Noise Only Mixed} & \textbf{1-Interferer Avg.} & \textbf{3-Interferer Avg.} \\
\midrule
$(1,\ +0)$ & 12.899 & 5.639 & 5.464 \\
$(3,\ +0)$ & \textbf{13.091} & \textbf{6.026} & \textbf{5.858} \\
$(5,\ +0)$ & 12.824 & 5.726 & 5.527 \\
\midrule
$(3,\ +2)$ & 12.984 & \underline{5.910} & \underline{5.643} \\
$(3,\ -2)$ & \underline{13.017} & 5.713 & 5.402 \\
\bottomrule
\end{tabular}
\vspace{-18pt}
\end{table}

\subsubsection{Impact of Graph Hyperparameters}
To examine sensitivity to graph hyperparameters, we vary the window size $k$ and the cross-modal shift $\delta$, with results shown in Table \ref{tab:ablation_hop}. 
We first observe the impact of window size with the shift fixed at $\delta=0$. 
Increasing $k$ from 1 to 3 yields gains, as a narrow window lacks sufficient receptive field to capture cross-modal temporal variation. 
However, expanding to $k=5$ degrades performance, as an oversized window introduces irrelevant distant frames and dilutes the graph's structural boundaries, risking negative transfer where acoustic interference erroneously aligns with distant visual cues.  
Regarding the cross-modal shift, both offsets ($\delta= \pm 2$) underperform the symmetry setting ($\delta=0$). 
This aligns with the nature of speech production: since audio-visual asynchrony is dynamic and sentence-dependent, forcing a unidirectional offset misaligns phonemes that require an alternate context.
The optimal performance of the symmetric, moderate window ($k=3, \delta=0$) suggests that our sparse graph topology functions as a localized tolerance buffer, encapsulating the dynamic range of asynchronies while bounding the propagation of cross-modal misalignment under adverse acoustic conditions.

\begin{table}[t]
\centering
\footnotesize
\caption{Cross-dataset generalization on VoxCeleb2.}
\vspace{-5pt}
\label{tab:vox2}
\setlength{\tabcolsep}{3pt}
\begin{tabular}{lccc}
\toprule
\textbf{Method} & \textbf{Noise Only Mixed} & \textbf{Interferer Avg.} & 
\textbf{Overall Avg.} \\
\midrule
RAVEN (Base) \cite{ma25c_interspeech} & 7.712 & 2.061 & 4.887 \\
Mamba & \underline{7.985} & \textbf{2.940} & \underline{5.463} \\
\midrule
SG-RAVEN & 7.945 & 2.514 & 5.230 \\
\textbf{SG-Mamba (Ours)} & \textbf{8.222} & \underline{2.911} & 
\textbf{5.567} \\
\bottomrule
\end{tabular}
\vspace{-18pt}
\end{table}

\subsection{Cross-Dataset Generalization} \label{cross}
Table \ref{tab:vox2} reports zero-shot transfer results on VoxCeleb2, where Interferer Avg.\ denotes the mean SI-SDR over 1- and 3-interferer conditions across 0, 5, and 10 dB SNRs, and Overall Avg.\ further includes the noise-only condition. 
SG-Mamba achieved the highest SI-SDR in the noise-only condition (8.222 dB) and the best Overall Avg.\ (5.567 dB). 
Under interferer conditions, Mamba achieved a marginally higher Interferer Avg.\ (2.940 dB vs.\ 2.911 dB), suggesting that the fixed graph topology becomes relatively less decisive under cross-dataset interference, consistent with the trend observed on LRS3. 
SG-RAVEN similarly improved over RAVEN across all conditions, suggesting that the structural benefits of the heterogeneous graph front-end transfer across datasets. 
These results demonstrate promising cross-dataset generalizability of the proposed framework without fine-tuning.

\section{Conclusion}
In this paper, we presented the SG-Mamba framework and showed that explicitly decoupling fine-grained cross-modal alignment from global temporal modeling helps address the ``Concatenation Trap'' in lightweight AVSE. 
Results on LRS3 and VoxCeleb2 show that explicit structural priors at the fusion stage can improve generalizability and robustness across diverse noise and multi-interferer conditions without substantially increasing computational cost. 
This graph-guided visual anchoring appears effective under severe acoustic interference. 
However, the current framework relies on a fixed graph topology and assumes uncorrupted visual inputs. 
Future work will investigate learnable sparse graph construction, robustness under audio-visual misalignment and visual degradations (e.g., occlusions), phase-aware reconstruction, and real-time streaming with end-to-end latency evaluation.

\section*{Acknowledgment}
This work was supported in part by Realtek. The views expressed do not necessarily reflect those of the sponsor.
We also used AI tools to assist with manuscript drafting and code development, with all outputs verified by the authors.


\bibliographystyle{IEEEtran}
\bibliography{references}

\end{document}